\documentclass[10pt,conference,final]{IEEEtran}
\usepackage{cite}
\usepackage[pdftex]{graphicx}
\usepackage{amssymb}
\usepackage[cmex10]{amsmath}
\usepackage[lined, ruled]{algorithm2e}
\usepackage{algpseudocode}
\usepackage{url}
\usepackage{enumitem}
\usepackage{multirow}
\usepackage{color}
\usepackage{subfigure}
\graphicspath{{./figure/}}

\newcommand{\ie}{{\em i.e.}}
\newcommand{\eg}{{\em e.g.}}
\newcommand{\etal}{{\em et al.\mbox{$\:$}}}

\newcommand{\name}{{Navigo }}
\newcommand{\nameNoSpace}{{Navigo}}

\begin{document}

\title{Navigo: Interest Forwarding by Geolocations in Vehicular Named Data Networking}

\author{
	\IEEEauthorblockN{Giulio Grassi\IEEEauthorrefmark{1}, Davide Pesavento\IEEEauthorrefmark{1}, Giovanni Pau\IEEEauthorrefmark{1}\IEEEauthorrefmark{2}, Lixia Zhang\IEEEauthorrefmark{2}, Serge Fdida\IEEEauthorrefmark{1} \\ 
	}
	\IEEEauthorblockA{
		\IEEEauthorrefmark{1} LIP6 -- Universit\'{e} Pierre et Marie Curie (UPMC), Sorbonne Universites -- Paris, France\\
		\{first\}.\{last\}@lip6.fr
	}
	\IEEEauthorblockA{
		\IEEEauthorrefmark{2} Computer Science Department -- University of California, Los Angeles, CA\\
		lixia@cs.ucla.edu
	}
}

\maketitle

\begin{abstract}
This paper proposes \nameNoSpace \footnote{This work has been submitted at WoWMoM 2015 in December 8th 2014}, a location based packet forwarding mechanism for vehicular Named Data Networking (NDN).
\name takes a radically new approach to address the challenges of frequent connectivity disruptions and sudden network changes in a vehicle network. Instead of forwarding packets to a specific moving car, Navigo aims to fetch specific pieces of data from multiple potential carriers of the data. The design provides (1) a mechanism to bind NDN data names to the producers' geographic area(s); (2) an algorithm to guide Interests towards data producers using a specialized shortest path over the road topology; and (3) an adaptive discovery and selection mechanism that can identify the best data source across multiple geographic areas, as well as quickly react to changes in the V2X network.
\end{abstract}
\vspace{-3.7 mm}

\section{Introduction}
\label{sec:intro}

Connectivity in a vehicular network is highly ad hoc. VANETs are made of dynamic, short lived links between vehicles that frequently come and go~\cite{rowstron2009characteristics}.  Traditional IP based node-to-node communication is a poor fit for the vehicular scenario because of the difficulties in coping with continuous disruptions and changes in network topology. 

Named Data Networking (NDN), in contrast, represents a radically different design.  An NDN network uses data names from applications directly for data delivery.  Because applications and their name spaces exist \emph{a priori}~\cite{VanSmThPlBriBra09-Networking}, NDN enables vehicles within vicinity of each other to exchange packets as soon as their signals can reach each other. Using application names eliminates the need for IP addresses.
However, in order to fetch data beyond the immediate neighborhood, vehicles need to make a decision on whether, and where, to forward Interests for specific content. 
In a wired network, this forwarding decision is assisted by NDN routing protocols which propagate data name prefixes throughout the network. In a vehicle network, data sources keep moving and connectivity changes frequently, making a routing protocol infeasible.

In this paper we describe \nameNoSpace, which exploits geographical information in place of a routing protocol to guide Interest forwarding in vehicular networks.  We develop solutions to (1) finding data sources' geographical information,
and (2) forwarding Interests along the best paths, taking into account specific properties of the vehicle networking environment.
We evaluate our design through simulations.

Our contributions can be summarized as follows.
First, we developed effective solutions to the problem of mapping data names to data locations and forwarding Interest packets along the best path.
Second, we designed an adaptive discovery and selection mechanism that can identify the available data sources across multiple geographic areas and can quickly react to sudden changes in vehicle networks. 
Third, our solution demonstrates the power of NDN architecture applied to vehicular networking and its ability to cope with ad hoc mobility and frequent network connectivity disruptions.

In the rest of the paper we first discuss the challenges in vehicular networking and identify the limitations in existing solutions in Section~\ref{sec:VANET-challenges.tex}.  We then sketch an overview of our solution in Section~\ref{sec:design}, and elaborate the design details in Sections~\ref{sec:forward} and~\ref{sec:LAL}).   Then we present the simulation scenarios and results in Section~\ref{sec:simulations}.  Lastly, we discuss and conclude the paper with some final remarks in Section~\ref{sec:discuss}.

\section{Vehicular Networks and Applications}
\label{sec:VANET-challenges.tex}
Vehicular networks are dynamic systems where nodes move frequently through the urban maze of roads and artifacts or they quickly pass by a stretch of high speed motorways. Node mobility and urban obstacles cause frequent connectivity disruptions and sudden network changes.  Previous work~\cite{rowstron2009characteristics} analyzed the morning rush hour vehicle-mobility of Portland, Oregon to understand the VANETs dynamics. In the Portland scenario, the link duration is less than 10 seconds in 97\% of the cases.
Mobility and urban artifacts are a challenge for the WiFi and DSRC connectivity in both vehicle-to-vehicle (V2V) and Vehicle-to-Infrastructure (V2I) cases.  The resulting network struggles to reach stability and frequently results in intermittent connectivity. 
~\cite{hadaller2007vehicular}~\cite{eriksson2008cabernet} show that 
the packet loss rate fluctuates due to inter-vehicle distance, interference, and presence of obstacles.  

\subsection{Existing Solutions and Limitations}
Vehicular networks have been studied for over a decade by industry and academia.  The literature includes both multi-hop routing and mobility management proposals, all based on IP's node-to-node communication model.
\cite{haerri2006performance}~\cite{jerbi2007improved} show that traditional ad-hoc networks routing does not perform well in VANET due to  short link duration and high protocol overhead.  
\cite{survey06}~\cite{li2007routing} proposed position-based routing protocols (PBR) that route packets toward  destination node positions rather than destination IP addresses.
In particular, GPSR~\cite{karp2000gpsr} performs stateless greedy routing towards the destination and handles failure using a graph planarization algorithm. It assumes all destination locations are known through querying an Oracle which supposedly has full knowledge of any node location in realtime.  GPSR maintains a 1-hop neighborhood state via a hello protocol;  and it uses unicast communications for packet forwarding. 
Short link durations and mobility force the protocol to frequently find new routes to moving targets and cope with the the combined effect of mobility and propagation \cite{giordano2010corner}.  Many recent PBR proposals try to learn from GPSR: introducing a DTN component~\cite{huang2008performance}, or using additional information such as car velocity, direction, and map awareness in order to adapt to changing vehicle density and scenarios~\cite{lee2008louvre}~\cite{lochert2003routing}~\cite{tian2009position}~\cite{naumov2007connectivity}.  While  \emph{second generation}  PBR's protocols show improved robustness and resilience in general, given they still attempt to route to a moving object in a maze of radio-opaque buildings, their performance pay a high price to mobility and high link volatility while battling to get the up-to-date scenario in order to properly adapt~\cite{lee2010survey}.
Within the Mobility Management proposals NEMO~\cite{devarapalli2005network} attempts to solve the IP mobility in the vehicular case by extending mobile IPv6. However, it has been shown that NEMO alone is insufficient in VANET and it requires a heavy infrastructure in the backend~\cite{zhu2011mobility}~\cite{baldessari2007nemo}. 

Navigo takes a radically new approach.  Instead of attempting to \emph{route to a moving object} in a complex urban scenario, Navigo aims to \emph{fetch a specific piece of data}  from a plethora of potential sources (e.g. producers, data mules, caches) by taking advantage of the broadcast nature of wireless medium, knowledge of digital maps, and NDN's pervasive caching.  It forwards data requests toward geographical regions where the requested data can be found.  This approach appears to be more robust to frequent topology changes and link breakages as success does not depend on the presence of any specific link, intermediate  node, or destination, as the requested data can be fetched from any node along the forwarding path or around the destination region.  This is a game changer that seamlessly integrates both V2V and V2I in a unifying VANET extension to NDN. 

\subsection{Named Data Networking}
The Named Data Networking (NDN) architecture was first introduced in \cite{VanSmThPlBriBra09-Networking}. An NDN network has three main players: data producers, consumers, and router/forwarders; they communicate by using application data names directly via two types of packets. A consumer sends an \emph{Interest} packet to request a specific piece of \emph{named data}, routers forward Interests toward data producers and keep track of all the pending Interests.
Each NDN node maintains three major data structures: Content Store (CS), Pending Interest Table (PIT), and Forwarding Information Base (FIB). The CS caches \emph{data packets} received, which can be potentially useful to satisfy future \emph{Interest} packets. The PIT stores \emph{Interests} that have been forwarded and waiting for matching \emph{Data} packets to return.  The FIB is similar to IP router's FIB and is maintained by a name-based routing protocol.  There is also a \emph{strategy module} that consults FIB in making Interest forwarding decisions.
 
\begin{figure}[thb]
\centering
\includegraphics[width=0.35\textwidth]{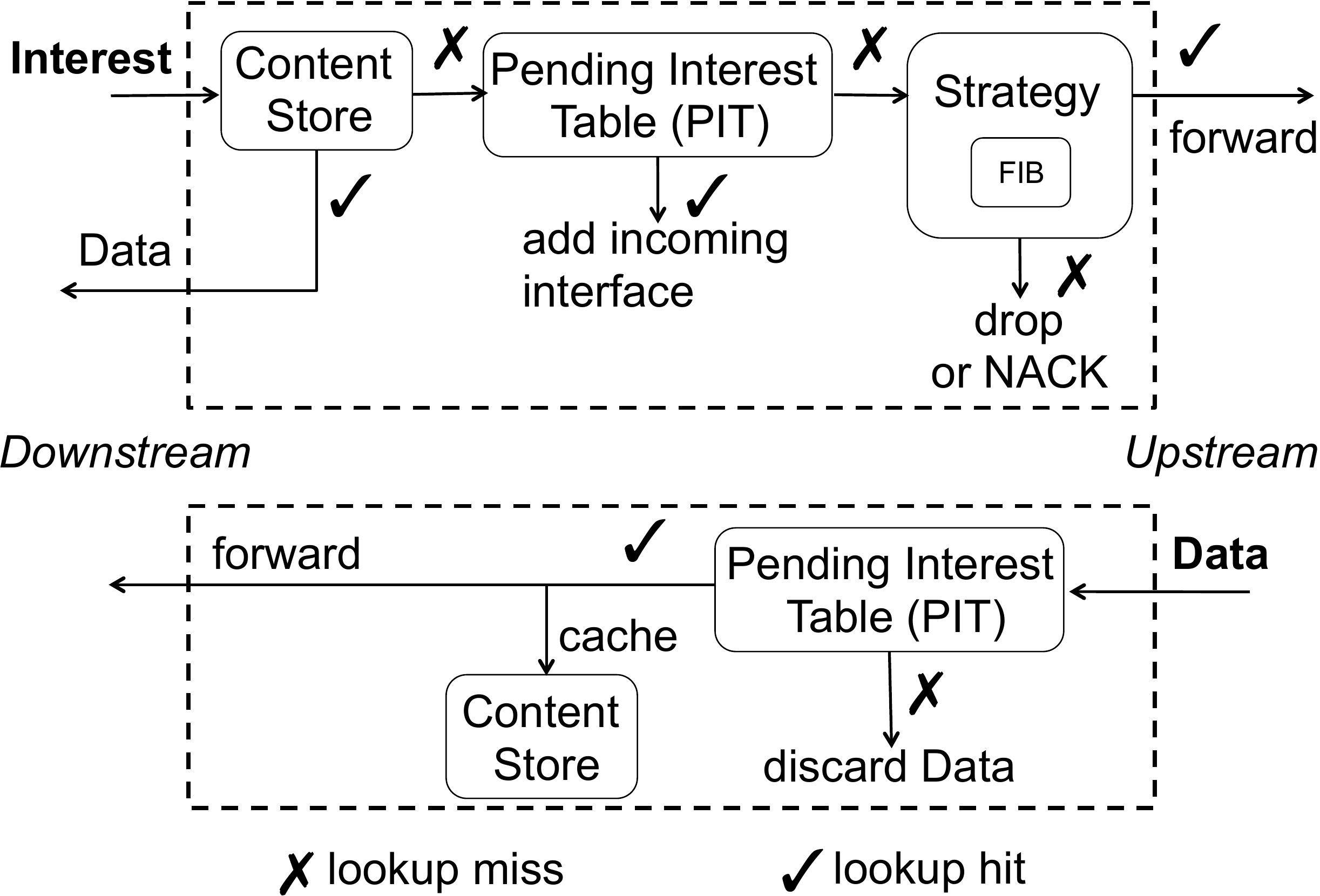}
\caption{Forwarding Process in an NDN node}
\label{fig:ndnforwarding}
\end{figure}

When an Interest meets a \emph{Data} packet with the matching name (either at the producer, or from a router cache), the Data packet follows the PIT entries left by the Interest to get back to the consumer. For each arriving Data packet, a router finds the entry in the PIT that matches the \emph{data name} and forwards the data to all downstream interfaces listed in the PIT entry. It then removes that PIT entry, and caches the Data in the CS.  See Figure~\ref{fig:ndnforwarding}.
Note that neither Interest nor Data packets carry IP addresses; Interests are forwarded toward data sources based on the names carried in them; Data packets return based on the PIT information set up by the Interests at each hop.

\subsection{Our Prior Work}
Our earlier work, dubbed V-NDN~\cite{2014NOM-VNDN}, shows the feasibility of applying NDN to vehicular communications.
V-NDN exploits the broadcast nature of the wireless channel to let all vehicles participate in communication, which helps overcome failures at any single link.  
Because IEEE 802.11 does not have collision detection/recovery for broadcast transmission, 
V-NDN introduces a Link Adaptation Layer (LAL) to amend this shortage.
First, LAL introduces the concept of implicit acknowledgment (ACK): After car $A$ broadcasts a packet $P$, if $A$ hears the transmission of $P$ by a neighbor (i.e. the packet being further forwarded), $A$ considers it an implicit partial ACK. 
$A$ stops retransmitting $P$ if it overhears implicit ACKs from each of the streets stemming from where the car is located.
Second, since $A$'s broadcast packet can be heard by multiple neighbors, to minimize collision and speed up packet propagation, neighbors set a random forwarding timer based on their positions: the farther one's distance from $A$, the smaller the waiting time before forwarding the packet. 
Third, when a node forwards a packet, it suppresses further transmission by all other cars between the previous hop and itself.

Although V-NDN proofs the feasibility and benefits of adopting the NDN paradigm for vehicular communication, it lacks a way to make smart forwarding decision, but blindly floods the Interest looking for the Content. Such an approach is not sustainable in a real deployment.
In this paper we design Navigo to steer Interest forwarding towards the data in order to build a viable V2V network.

\section{Navigo Design Overview}
\label{sec:design}

A fundamental challenge in Navigo design is how to steer Interest towards where data resides. As we discussed in Section~\ref{sec:VANET-challenges.tex}, the highly dynamic connectivity in VANET renders running a routing protocol infeasible.
For traffic applications that intrinsically contain location info in data names, \cite{wang2012rapid} demonstrated that one can simply forward Interests toward the geographic location stated in the names, without the need for a routing protocol. Indeed a broad class of automotive applications is intrinsically location-dependent, i.e. the data produced and consumed by them is tied to specific locations. Examples of such applications range from obtaining road traffic updates on a given street to the search for an available parking space. However, \cite{wang2012rapid} requires the forwarding strategy in each node to understand the semantic of the names to be able to extract the destination information. Such assumption ties the name design with a specific convention,  and most important, is not feasible in the current NDN framework, where the forwarding strategy is unaware of the name semantic. A means of letting consumers suggest where the data may reside is still missing. Furthermore, other types of applications, such as music sharing or data fetching in general, are not associated with any specific locations.


To effectively forward Interests for all types of applications without a routing protocol, our solution is to couple their data names with the locations of where the data resides. While for the first type of application we can bound names with the location the consumer is interested in, for the second type our solution is to bind them with the location of the data provider: either the content producer or a vehicle which is carrying the data in its cache (mule) or an Internet access point (i.e. RSU). This will allow us to do geo forwarding to support all types of applications.

There are a number of specific issues to address to make the above idea work. First, one must define a namespace for geo locations; this is addressed below. Second, one must provide effective means to map data names to locations, which is discussed in Section~\ref{sec:bindingNamesToArea}.
Third, we would like to support geo forwarding with no modification to the existing NDN forwarding framework; this is addressed in Section~\ref{sec:binding}.

\subsection{Naming geographic areas}
\label{sec:geoarea}

\begin{figure}[htb]
\begin{minipage}[htdp]{\columnwidth}
\vspace{-0.5cm}
\centering{
\subfigure[Example of MGRS map]{
\includegraphics[width=0.36\columnwidth]{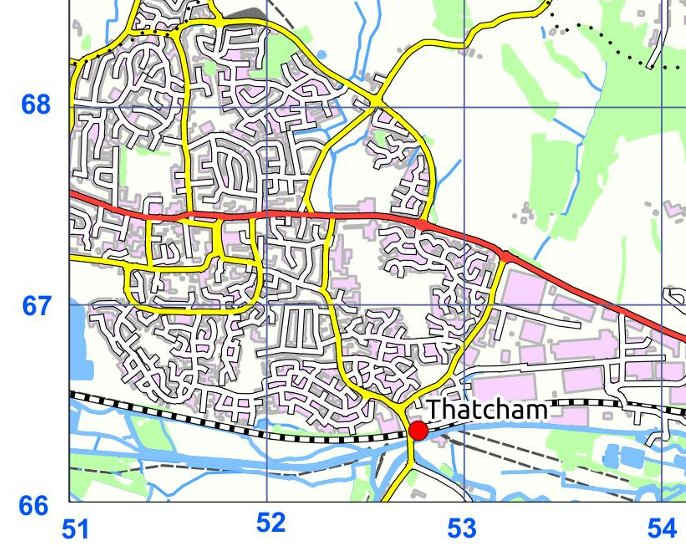}
\label{fig:mapgrid}
}
\hfil
\subfigure[Mapping GeoFaces to geographic areas]{
\includegraphics[width=0.56\columnwidth]{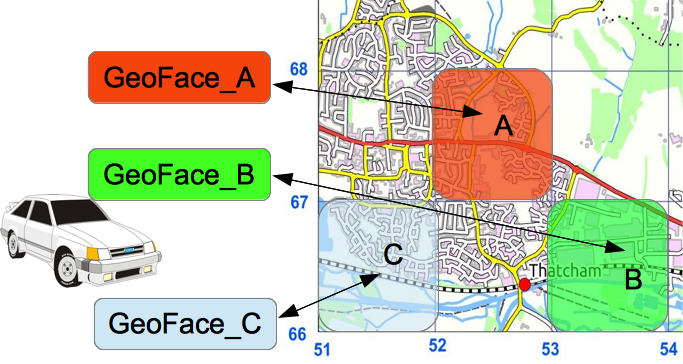}
\label{fig:GeoFace}
}
}
\vspace{-0.5cm}
\caption{}
\label{fig:MobilityPatterns}
\end{minipage}
\end{figure}

We divide the world into regions according to the Military Grid Reference System (MGRS). This system is derived from the Universal Transverse Mercator (UTM) and from the Universal Polar Stereographic (UPS) grid systems, where each region is identified by a label (Figure~\ref{fig:mapgrid}).

In the current implementation geo-areas have a fixed size (200x200 meters) but the MGRS scheme easily allows names with different precision levels, by adding or removing digits from the name, \eg{} 4QFJ 12 67 defines a 1Km precision, while QFJ 123 678  has a 100 meter precision. Further analysis regarding more flexible areas is left for future research.


\subsection{Mapping data names to geo-areas}
\label{sec:bindingNamesToArea}
When a node has an Interest in hand for data without any geographic meaning and has no knowledge about the prefix of the data name (not in its FIB), the node simply sends the Interest out in all directions.
If any of these flooded Interests hits a copy of the matching data, the responder attaches its geo-area MGRS name (\eg{} 4QFJ 123 678) to the returned Data packet. As the Data packet follows the breadcrumb trace of the Interest, all the nodes along the way learn the binding between the data name prefix and the corresponding geo-area. This information allows them to forward future Interests for the same name prefix to that geo-area only.
For location-dependent data, the consumer can avoid the initial Interest flooding procedure by binding the Data name with the geo-area the consumer is interested in before sending the Interest. 

\subsection{Hiding geographic forwarding from basic NDN framework}
\label{sec:binding}

The current NDN forwarding daemon has no concept about geo-areas. Indeed its FIB contains $\langle$prefix, face$\rangle$ pairs only. To exploit the binding among names and geo-areas while using the current NDN architecture \name introduces the concept of ``Geographic Faces'' (GeoFaces), and implements the binding of name prefixes to geo-areas through a two-step process.
It first binds a geo-area to a GeoFace, and then lets the FIB store the mapping from the name prefix to the GeoFace as in the current NDN architecture.
One (\ie{} the consumer) can now simply register a new rule in the FIB to bind a prefix with a specific geo-area.

An extended version of the LAL, introduced by~\cite{2014NOM-VNDN}, stores the GeoFace to geo-area mapping. Such GeoFaces are an abstraction of the WiFi ad hoc interfaces the car is equipped with. \name extends the LAL by implementing a forwarding mechanism which, given a GeoFace $F_X$ bound to a geo-area $X$, steers the Interest over the V2V channel along the best path to $X$, according to Section~\ref{sec:LAL}.
Furthermore, the LAL sorts incoming Data packets and redirects them to the correct GeoFace based on the Data provider's location.

\begin{figure}
\vspace{-0.4cm}
\begin{minipage}[t!h]{\columnwidth}
\centering
\includegraphics[width=0.75\columnwidth]{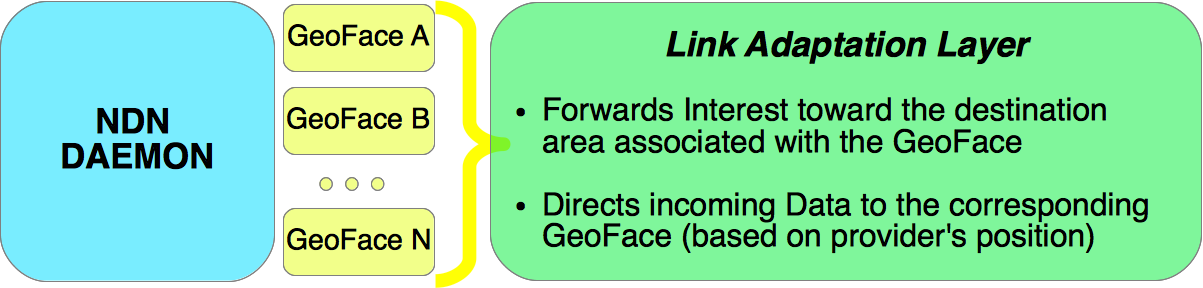}
\vspace{-0.3cm}
\caption{NDN and GeoFaces}
\label{fig:GeoFaceArch}
\vspace{-0.5cm}
\end{minipage}
\end{figure}


\subsection{Design assumptions}
This paper is focused on urban scenarios. We assume all vehicles are equipped with a GPS sensor and a digital map, therefore able to identify their own location (and thus its MGRS name). Each vehicle is also equipped with a WiFi Ad-Hoc communication based on WiFi and with enough storage and computational capabilities; we leave the study of the performance under constrained caching to a future work.
Furthermore, we assume the presence of RSUs along some roads which provide Internet connectivity to all the cars and which are running a NDN stack and \nameNoSpace.


\section{Geolocation-based Interest Forwarding}
\label{sec:forward}

The main idea behind the forwarding strategy adopted by \name is to explore the area surrounding the node looking for producer, mules or RSU and then, as soon as the first Data packet comes back, forwarding future Interests for the same prefix towards the geo-area where the data is coming from.

\subsection{Forwarding algorithm}
\label{sec:ForwardingAlg}

When forwarding an Interest, \name tries to exploit the presence of multiple data providers to balance traffic and make communication more resilient to the high dynamism of VANET, which limits the validity in time of a single rule.
To forward an Interest \emph{I} for a Content named \emph{N} \name adopts the following forwarding strategy:

\begin{itemize}
	\item If \emph{I} does not match any entries in the FIB (there is no information about the Data location), the Interest is sent using a flooding technique (see \ref{sec:LAL}) without specifying any destination area---the node is in a so-called \emph{exploration phase}.
	\item Instead, if several GeoFaces are bound with \emph{N}, the forwarding strategy selects one face in a round-robin way.
	\item If only one face is available, with probability \emph{p} (0.95 in our experiments) the GeoFace in the FIB is used, while with probability $1-p$ the node acts as in the \emph{exploration phase}, flooding the Interest. \name adopts this additional exploration phase to avoid focusing on a single destination area for a long time while some other opportunities may come with mobility to balance the traffic.
	\item After sending an Interest, if the node does not receive the requested data before a deadline T (300ms in our experiments), the binding in the FIB among the face and \emph{N} is removed. 
	\item If multiple Interests with the same prefix \emph{N} are sent in pipeline on the same GeoFace \emph{F$_{X}$}, whenever one of them is satisfied, the deadline is removed for all the pending Interest for \emph{N}: receiving a Data packet indeed means the information is still available in that geo-area, although some of the Interest may fail. 
\end{itemize}

An evaluation of using more sophisticated criteria, such as round trip time or success rate, to select the best face is left for future works.

\section{FIB Management}
\label{sec:FibManagment}

\subsection{Binding content location to the right name}
Names in NDN are hierarchical: a Content named \emph{/N} may be divided in several pieces i.e. \emph{/N/c1}; \emph{/N/c2}. To correctly forward Interest for all these pieces (and only them), one must have a FIB entry with the prefix \emph{/N} (FIB lookup is done by longest prefix match). Consumer and producer know the name semantic and thus can identify which prefix aggregates all the pieces of a Content. But forwarders, which may not be running the application, are unaware of such thing and thus they cannot correctly fill the FIB\footnote{Assuming that by removing the last component of the name one gets the correct prefix to aggregate all the Interests for the Data may not always be true.}. To address this issue, the consumer's LAL attaches to each Interest (into a 2.5 layer header defined by LAL) the prefix which aggregate all the pieces of the requested content. This information is then spread by the forwarders, allowing every LAL to register the correct rule in the FIB whenever the Interest is satisfied.


\subsection{FIB size}
A FIB entry stores a prefix and a list of faces that can be used to retrieve such Content. Associating geographic areas with faces, the geo-area dimension may affect the size of the FIB: smaller areas means higher probability to receives Data from different region, especially when the data provider is a moving car, which increases the number of faces bound with the same prefix. At the same time, however, if the region is too large, the FIB is smaller but the portion of area where cars flood the Interest increases, leading to a larger overhead. 

With 200$\times$200 meter areas our experiments show that the number of faces bound with the same prefix never reaches values higher than 9. Indeed, removing the binding among a GeoFace and a prefix as soon as an Interest fails, stops the node from having too many faces bound to the same prefix.

%
%

\section{Link Adaptation Layer}
\label{sec:LAL}

\name extends the original version of the LAL presented in~\cite{2014NOM-VNDN} exploiting the knowledge of the destination area. It takes care of the GeoFaces to geo-areas binding, it steers Interest along the shortest path to the destination region and it copes with urban scenario characteristics by taking into account the presence of obstruction in contrast to more suitable places for wireless propagation.


\subsection{LAL and GeoFaces}
\label{subsec:LALandGeoFaces}
The LAL creates and destroys GeoFaces and keeps the mapping between GeoFaces and geo-areas in the Face-to-Area table (F2A). Whenever a node receives a Data packet, the LAL extracts the geo-areas information attached by the data provider. If this area is not associated with any faces, the LAL creates a new GeoFace and binds it to the geo-area (adding the relation to the F2A). Whenever a GeoFace is unused for a certain amount of time (\ie{} order of tens of seconds), the LAL removes the face and any references to it from the F2A and the FIB.

\subsection{Calculating the shortest path}

The shortest path to the destination area is calculated applying a specialized Dijkstra algorithm with the street topology as underneath graph: streets are edges and intersection are nodes of the graph. Conceptually \name deliberately substitutes the more stable road-topology to the network topology which in VANETs is dynamically partitioning and features short lived links lasting only few seconds on average~\cite{rowstron2009characteristics}. In computing the shortest path, LAL is unaware of neighbors locations. Indeed, to avoid the overhead of periodic exchanges of 1-hop neighbor position\cite{HelloprotocolsOverhead}, \name doesn't use any neighboring protocol. It relies on a probabilistic approach to minimize the chances of hitting an empty road. \name assigns costs to edges that are inversely proportional to the number of lanes\footnote{After preliminary simulation, we adopted the following costs: 1 for 2-lanes road, 0.7 for 4-lane street and 0.25 for 6-lanes roads. Analysis of other factors, such as amount of data traffic or cars to determinate the weights of a road are left for future works.}. This way the algorithm tends to prefers paths with larger roads and more likely to have running cars at any moment thus leading to a more stable path. Furthermore, the algorithm takes into account the presence obstruction of wireless communication and merges roads that are in line of sight while splits paths that require a turn. Indeed turns mean additional hops in the transmission, which increases overhead and delay.

\subsection{Forwarding process}
Whenever the NDN daemon sends an Interest through a GeoFace, the face passes the packet to the LAL, which perform a lookup on the F2A to determines the destination area name. This information is then encoded within the L2.5 header that encapsulates the Interest, together with the position of the node, spreading the information to all the neighbors in the transmission range. Once a car receives an Interest, the LAL extracts and stores the information about the destination area, the position of the previous node and the nonce of the Interest and then it passes the packet to the forwarding strategy. If the NDN daemon decides to forward the Interest on the V2V network, it passes the packet back to the LAL, either by using the GeoFace specified in the FIB\footnote{If necessary, \name allows the outgoing and incoming face to be the same} or by using the exploration phase procedure (see the following paragraph on Flooding). The LAL, based on the nonce of the Interest, recovers the position of the previous hop and the destination area specified by the original consumer, which is used as the Interest destination. LAL overrides any local decision about the destination area with the consumer's will. The analysis of benefits and challenges of overriding the consumer's will with local information is left for future works.

Given the destination area, LAL computes the shortest path algorithm and forwards the Interest only if it's closer (path is cheaper) to the destination area than the previous hop. 
Once the Interest reaches the destination area \name uses the protocol described in~\cite{2014NOM-VNDN} to perform a local Interest dissemination, flooding the Interest in all the available directions. Cars outside the destination area may reply with the Data, but they don't forward the Interest anymore, constraining its dissemination to the destination area only. Whenever the Interest hits a data provider, the node replies with the Data, attaching its location MGRS name \emph{X} to the 2.5 layer header. 
As defined by the NDN protocol, Data will follow the breadcrumbs left by the Interest in the PIT of every nodes it passed through. In this way back to the consumer, LAL updates the FIB binding the Content prefix with the GeoFace associated with \emph{X} and forwards the Data only if it is closer than the previous hop to the node from which it received the Interest.

To increase the communication reliability, during the packet forwarding process LAL utilizes the implicit acknowledgment concept introduced by~\cite{2014NOM-VNDN}, requiring an implicit ACK from the street indicated by the shortest path as next hop.\\

\subsubsection*{Flooding}
\label{sec:flooding}
When the forwarding strategy is in \emph{exploration phase}, it selects the v2vFace as outgoing face for the Interest. Such v2vFace, introduced by~\cite{2014NOM-VNDN}, is used to spread the Interest over the V2V channel in all the directions. The destination area is not specified and the LAL adopts the packet suppression techniques defined in~\cite{2014NOM-VNDN} to flood the network.


\begin{algorithm}
\caption{LAL -- Interest Forwarding}
\label{alg:LALForwarding}
\small
\tcc{LAL receives an Interest}
\KwData{Interest \emph{I}.}
{
	$Nonce \longleftarrow$ ExtractNonce($I$)\;
	Extract $PHPos$ (previous hop position) from 2.5 header\;
	Extract $DA$ (destination area) from 2.5 header\;
	$GeoFace \longleftarrow$ Lookup($F2A, DA$)\;
	\tcc{Create GeoFace for DA if needed}
	
	Store $Nonce, PHPos, DA$ in $InterestFromNetwork$\;
	Pass \emph{I} to forwarding strategy using $GeoFace$\;
	\tcc{If DA is not specified by the consumer (exploration phase), use v2vFace}
}

***********************************************

\tcc{LAL receives an Interest from the forwarding strategy}
\KwData{Interest \emph{I}; my position \emph{MyPos}; face used \emph{F}.}
{
	$Nonce \longleftarrow$ ExtractNonce($I$)\;
	
	\eIf{$Nonce \in InterestFromNetwork $}{	
	$Distance \longleftarrow$ CalculateDistance($MyPos, PHPos$)\;
	\eIf{$DA$ defined for $I$}{
		\tcc{Calculate path cost}
		$PrevHopCost \longleftarrow$ Dijkstra($PHPos,DA$)\;
		$\langle NextHop, Cost\rangle \longleftarrow$ Dijkstra($MyPos,DA$)\;
		\eIf{$Cost < PrevHopCost$} {
			CalculateWaitingTimer($Distance, MyPos$)\;
			AttachToPacket($I,MyPos,DA$)\;
			Send($I$)\;
			WaitForAckFrom($NextHop$)\;
		}
		{
			Stop processing \emph{I}\;
		}
		}
		{
		\tcc{No DA specified in \emph{I}}
			CalculateWaitingTimer($Distance, MyPos$)\;
			Send($I$)\;
			WaitForAckFrom($AllPossibleDirections$)\;
		}
	}
	{
	\tcc{\emph{I} generated locally}
		\eIf{$F \in GeoFaces$}{
			$DA \longleftarrow$ GetCoordinates($F$)\;
			$\langle NextHop, \_\rangle \longleftarrow$ Dijkstra($MyPos,DA$)\;
			CalculateWaitingTimer($Distance, MyPos$);\
			AttachToPacket($I,MyPos,DA$)\;
			Send($I$)\;
			WaitForAckFrom($NextHop$)\;
		}
		{
		\tcc{Exploration phase on v2vFace}
			Send($I$)\;
			WaitForAckFrom($AllPossibleDirections$)\;
		}
	}
}
\end{algorithm}

\subsection{Forwarding based on forwarding points}
\begin{figure}
\begin{minipage}[t!h]{\columnwidth}
\vspace{-0.2cm}
\centering
\includegraphics[width=0.3\columnwidth]{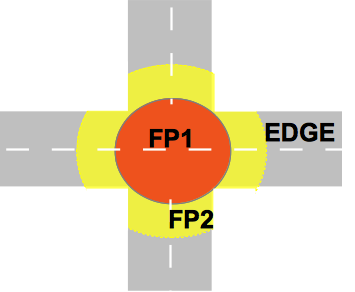}
\vspace{-0.3cm}
\caption{Intersection as forwarding points.}
\label{fig:forwardingPoint}
\vspace{-0.4cm}
\end{minipage}
\end{figure}

As introduced in \cite{frank2010trafroute}, due to the presence of buildings that obstruct wireless communication, the best strategy to cover a large area with the smallest number of hops is selecting forwarders at the intersections. 
\name capitalizes this observation and uses the junctions as preferred forwarding points, by speeding-up the transmission of cars within an intersection. As \cite{frank2010trafroute}, \name splits every junction into two parts: the core, so-called FP1, and the external area, so-called FP2 (see figure~\ref{fig:forwardingPoint}). Among vehicles inside the same junction, LAL privileges cars within FP1, which have a more central position, increasing the chances to reach more cars in one shot. \\

When forwarding a packet, to maximize the progress at each hop the LAL in~\cite{2014NOM-VNDN} assigned higher priority to nodes far from the sender by speeding up their transmissions.
\name extends this approach, by taking into account not only the distance by the previous hop, but also the presence of forwarding points and the type of packet:
\begin{itemize}
	\item Vehicles inside a FP wait less. A random component is added to the waiting timer to avoid collision among cars at the same junction. Furthermore, cars inside FP1 waits less than vehicles in FP2.
	\item Among cars located in different FPs, the shorter the distance to the previous hop, the longer the wait: the road among the current and the previous hop is divided in 100 meters sections. Cars in the FP within the furthest sector from the previous hop (distance greater than 500 meters) waits for the minimum waiting timer. Getting closer to the previous hop, each 100 meters segment adds a constant value (4 ms for Data, 1.5 ms for Interest) to the waiting timer. 
	\item The waiting timer for cars at an edge is inversely proportional to the distance to the previous hop (as in V-NDN).
	\item The maximum waiting timer for a Data packet is smaller than the minimum waiting timer for an Interest. Speeding up the transmission of Data stops the data provider neighbors to propagate the Interest any further. 
\end{itemize}
The entire process of calculating the waiting timer is self-deterministic: each car calculates its own waiting timer based on its position and the distance to the previous hop, without requiring any knowledge about neighbors position.

\begin{figure}
\begin{minipage}[t!h]{\columnwidth}
\centering
\vspace{-0.2cm}
\includegraphics[width=0.9\columnwidth]{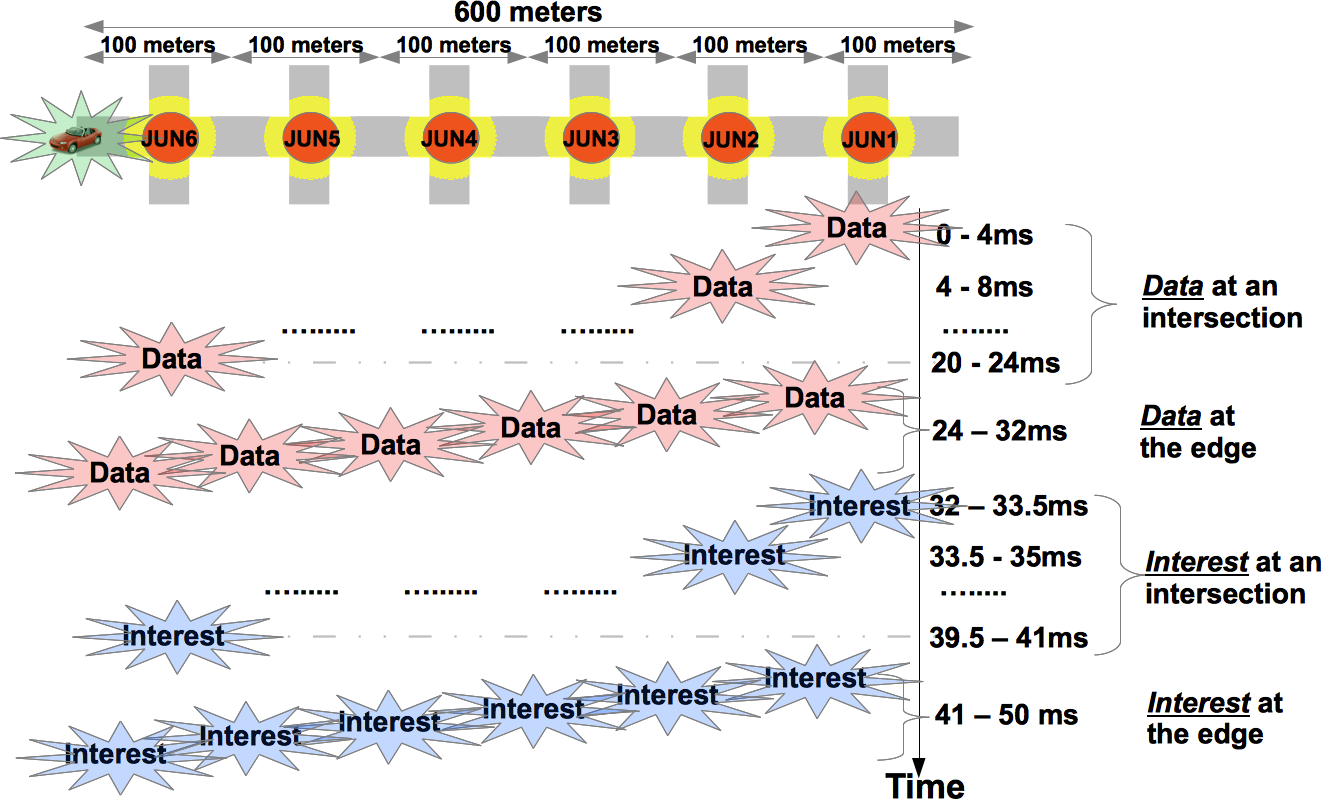}
\vspace{-0.3cm}
\caption{\name speeds up Data transmission and prioritizes packet forwarding at the intersections. Cars close to the previous hop and cars not located within a forwarding point that are waiting to send the same packet will suppress their transmission.}
\label{fig:forwardingByFP}
\vspace{-0.4cm}
\end{minipage}
\end{figure}

Figure \ref{fig:forwardingByFP} shows how much a car waits before forwarding a packet\footnote{The values shown in Figure~\ref{fig:forwardingByFP} correspond to the values used in the experiments. While for simplicity these values are constant, adapting the waiting timer based on the environment (\eg{} car density, data traffic, \dots) can improve the performance.}.
It must be noted that even though it might take at most 50 ms to make a one-hop progress for a packet, the delays sensitively reduce when a Data packet has to be forwarded or when the car trying to forward an Interest is either at an intersection or far from the previous hop.
Furthermore, based on the implicit acknowledgement policy already discussed, as soon as one car forwards a packet, all the vehicles with larger waiting timers suppress their transmission.


\section{Simulations}
\label{sec:simulations}

\subsection{Scenario}
\label{sec:simscenario}

For the initial evaluation of our design we considered an urban road network with both residential streets and major arteries.
The map we chose for the simulations spans a $2.1 \times 2.1$ km area in the city of Los Angeles, CA.
The vehicular micro-mobility traces were generated with SUMO~\cite{SUMO2012}.
In order to make the simulated scenario as close to reality as possible, the traffic volume is shaped according to the importance and number of lanes of each street: out of 812 cars, 48\% of them were on 6-lane roads, 37\% on 4-lane roads, and the remaining 15\% was on 2-lane residential streets.
The average time a vehicle spends inside the simulated area is 3 minutes.


For simplicity, every car enters the maps with an empty Content Store, even though this configuration penalizes \nameNoSpace, which heavily relies on nodes' caches.
Moreover, similarly to \cite{2014NOM-VNDN}, we observed that deploying large storage devices in cars should not be a problem nowadays, and therefore we set the Content Store size limit to 10 GB on each car, which means the car can keep the received data in the cache for the entire duration of its trip.

All vehicles are equipped with an IEEE 802.11 wireless network interface operating at 24 Mbit/s, configured in ad hoc (IBSS) mode on the same fixed channel used by the roadside units (RSUs), thus allowing cars to communicate with both other vehicles and RSUs via the same interface. 

The roadside units are positioned to mimic the location of the actual access points deployed by Time Warner Cable in the same area. We selected only a subset of 4 out of the 21 access points currently deployed in that area~\cite{TWCwebsite}. In our model RSUs are assumed to be fully functional NDN nodes; furthermore, we assume that no MAC-layer authentication or link setup process is needed. Although this does not reflect current WiFi practices, emerging standards such as the vehicular-specific IEEE \mbox{802.11p} and the proposed IEEE \mbox{802.11ai} either do not require link-layer connection establishment, or they reduce the link setup time to less than 100 milliseconds.

We implemented \name on top of a modified version of ndnSIM, an ns3-based NDN simulator~\cite{ndnSIM}.
The radio signal propagation was modeled with CORNER~\cite{giordano2010corner}, a high-fidelity propagation model for urban scenarios that accounts for the presence of buildings as well as fast-fading effects.



\subsection{Music streaming over NDN}
\label{sec:apps}

In order to evaluate \name performances, we devised a ``music streaming'' application: a hypothetical Internet music streaming provider that can be reached by any of the 4 RSUs deployed on the map via a 100 Mbit/s wired channel.
The client application (consumer) ran on a subset of all the cars; we varied the cardinality of the subset from 2\% to 100\% across our simulations.

Each song has an average length of 3 minutes, yielding about 1700 chunks of data per song, if we assume an average encoding bit rate that is common among current commercial music streaming services such as Spotify.
Requests for songs are generated according to a Zipf distribution with an $\alpha$ parameter obtained from \cite{kreitz2010spotify}, where Kreitz \etal found that the top 12\% most popular songs in the library are requested 88\% of the time.
When a song is chosen, the consumer starts issuing Interest packets progressively for every chunk of which the song is composed.
To improve the performance we implemented a simple mechanism for request pipelining, with a hard limit of 20 pending Interests (\ie{} expressed but not satisfied) at any given time.

The application tries to provide the best possible user experience, thus its main goal is to successfully retrieve a chunk before the playback reaches that point of the song.
In order to do so, and since Data packets can arrive out-of-order, the streaming client maintains a buffer of song fragments that have already been fetched but have not yet been played.
When this buffer underflows, the application has to pause the playback and wait for the missing chunk before playback can be resumed.
This event is highly undesirable since it leads to a poor user experience.
In our simulations we recorded whether an underflow occurred during the playback of a song.
We believe this metric provides an important tool to evaluate the success of our solution.

\subsection{Simulation results}
\label{sec:simresults}

We compared \name to GPSR~\cite{karp2000gpsr}, a well-known routing protocol for mobile wireless networks that uses the geographic positions of nodes to make packet forwarding decisions. GPSR typically requires a location service to discover the data source position: in our experiment we provided GPSR with a cost free oracle able to locate the closest node (server or consumer) with the requested chunk of the song.
In this section we present the results obtained from the simulations and we analyze them.

\subsubsection{Success rate}
\begin{figure*}
\begin{minipage}{\textwidth}
\vspace{-0.2cm}
\subfigure[Success rate]{
\includegraphics[width=0.23\columnwidth]{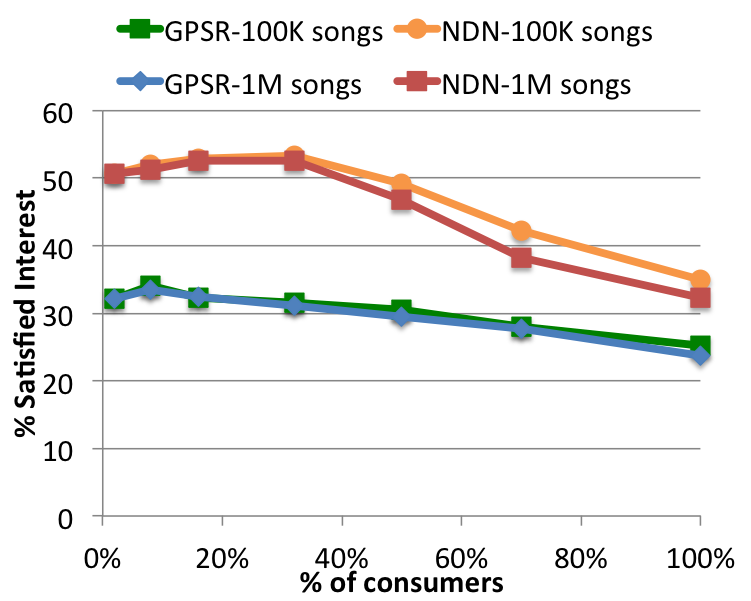}
\label{fig:SuccessByLibrarySize}
}
\hfill
\subfigure[User satisfaction]{
\includegraphics[width=0.23\columnwidth]{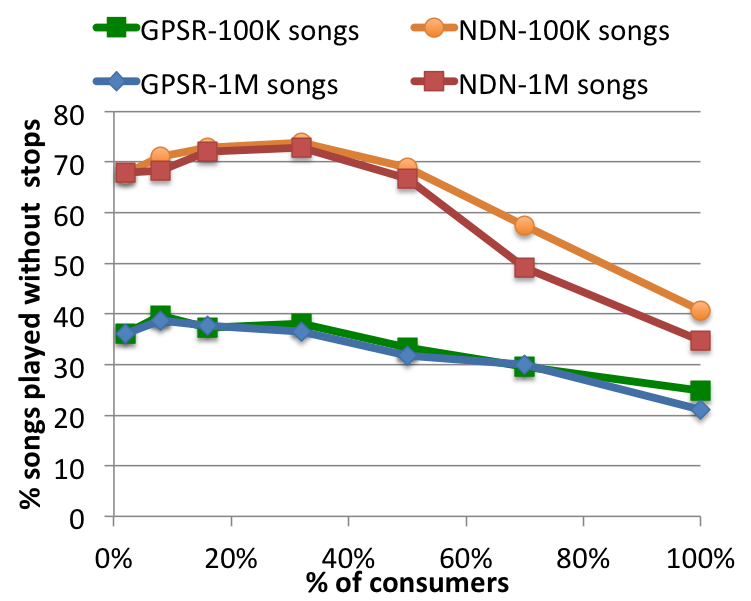}
\label{fig:PlayedWithoutInterruptionByLibrarySize}
}
\hfill
\subfigure[Protocol overhead]{
\includegraphics[width=0.23\columnwidth]{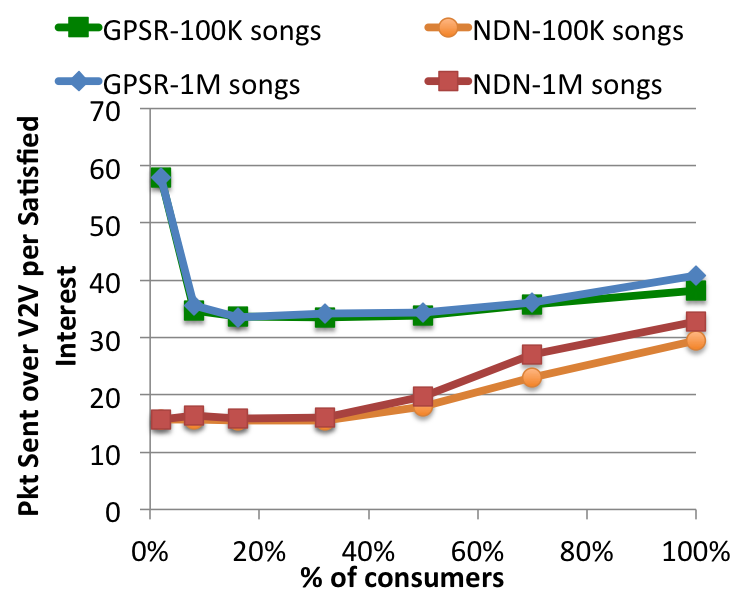}
\label{fig:ChannelAccess}
}
\hfill
\subfigure[Load on the infrastructure]{
\includegraphics[width=0.23\columnwidth]{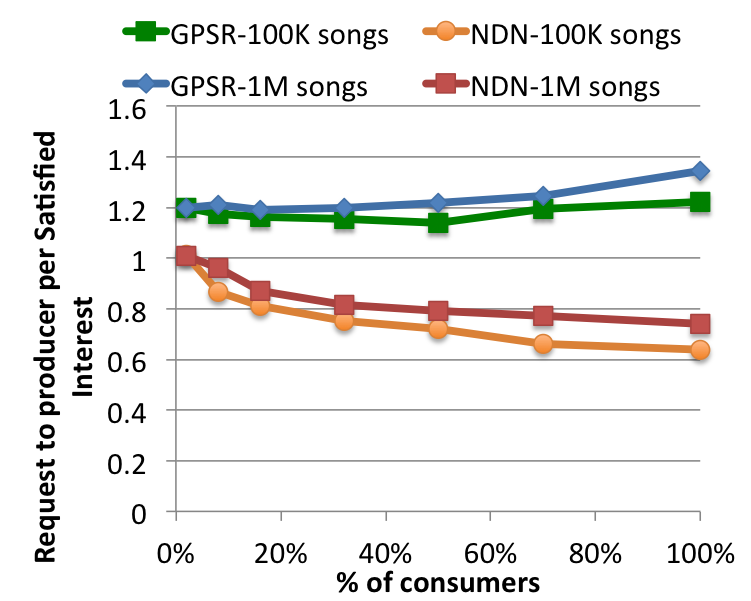}
\label{fig:RequestToProducer}
}
\end{minipage}
\vspace{-4mm}
\caption{Performance with different music library sizes}
\vspace{-0.5cm}
\end{figure*}

Defined as the ratio between the number of satisfied Interests and the number of Interests issued by all consumers.
The results for this metric are shown in Figure~\ref{fig:SuccessByLibrarySize}.
\name is able to satisfy a much higher percentage of Interests compared to GPSR, and although the margin of improvement shrinks with 70\% and 100\% of consumers, \name can still satisfy 10\% more Interests than GPSR.

\subsubsection{User satisfaction}
As explained in \ref{sec:apps}, this metric is expressed as the percentage of songs played without interruptions caused by a playback buffer underrun.
During our simulations the maximum buffer size was set to 30 seconds.
Figure~\ref{fig:PlayedWithoutInterruptionByLibrarySize} shows that \name substantially outperforms GPSR, especially with 50\% consumers or less.


\subsubsection{V2V channel access (protocol overhead)}
Represents the intrinsic ``cost'' of the V2V protocol in terms of number of accesses to the WiFi channel needed to satisfy an Interest.
For \name this means the average number of Interest and Data packets that were sent on the air for each satisfied Interest.
For GPSR we counted also ARP, ICMP, and Hello packets, which are required to run the protocol and therefore part of its overhead.
We can see in Figure~\ref{fig:ChannelAccess} that GPSR requires a much higher number of accesses to the V2V channel compared to our solution.
In particular, while GPSR always needs to send more than 30 packets for each satisfied Interest, \name requires less than 20 packets in most cases, increasing the overall network efficiency, and only becomes slightly worse with more than 50\% of consumers.


\subsubsection{Load on the infrastructure}
Expressed as the number of requests received by the streaming server (located on the Internet behind the RSUs) divided by the total number of Interest satisfied.
This metric is particularly interesting as it illustrates a major limitation of IP's approach.
Indeed, with IP-based protocols, it sometimes happens that the request reaches the content provider but the response fails to travel back to the consumer.
In this case the consumer has to re-request the data from the content provider, because there are no caches along the path and, from the point of view of IP, the two requests are completely different and unrelated, even if they refer to the same content.
This of course does not happen with NDN: Interests re-issued after a timeout can be satisfied by any other node that cached the desired Data packet during the previous failed attempt(s).
The effect is evident from Figure~\ref{fig:RequestToProducer}.
The inability of GPSR to exploit in-network caching results in a load ratio around 1.2 or higher in all scenarios.
On the other hand \name never goes above 1, and in most cases the load is around 0.8.
Moreover, as the number of consumers grows, our solution is able to take advantage of the increased caching opportunities, thus lowering the load on the infrastructure even more, contrary to GPSR where the load slightly increases.


\subsubsection{Infrastructure offload}
\begin{figure*}
\begin{minipage}{\textwidth}
\vspace{-0.2cm}
\subfigure[Infrastructure offload]{
\includegraphics[width=0.31\columnwidth]{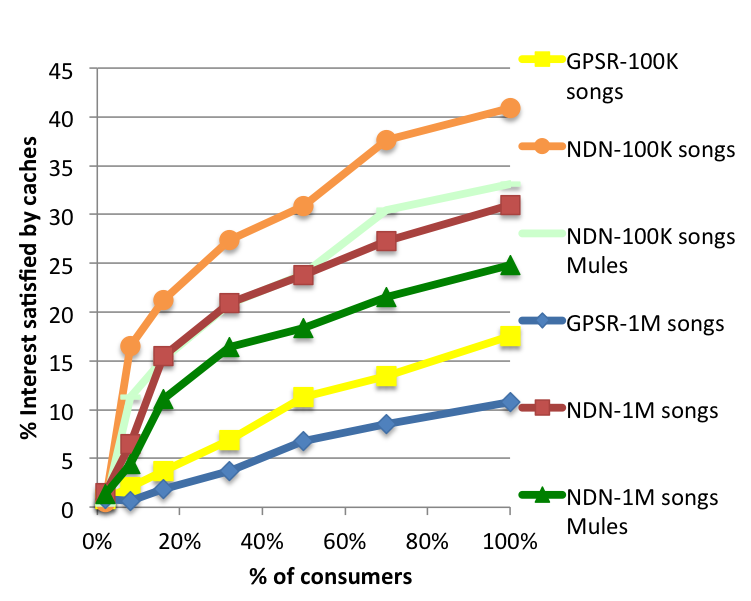}
\label{fig:InfrastructureOffload}
}
\hfill
\subfigure[Consumers opportunistically retrieve a song from different mules]{
\includegraphics[width=0.31\columnwidth]{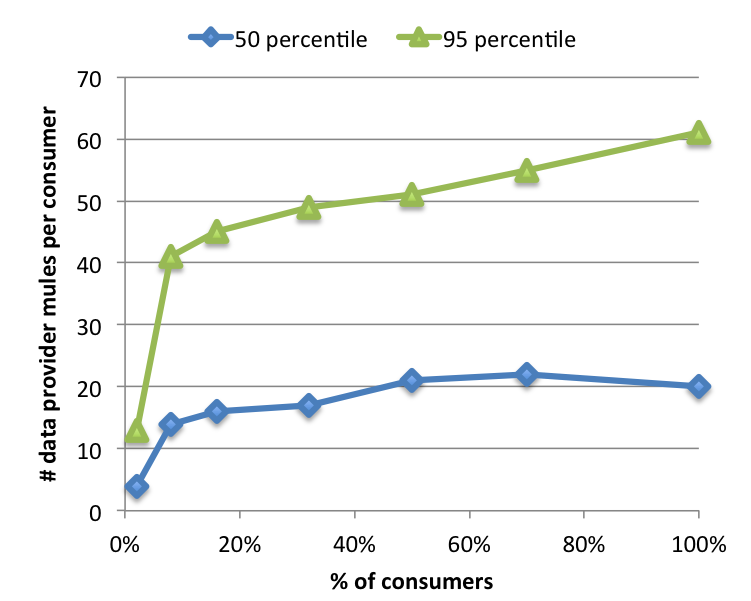}
\label{fig:MulesPerConsumer}
}
\hfill
\subfigure[Transmission queue length with different car densities]{
\includegraphics[width=0.31\columnwidth]{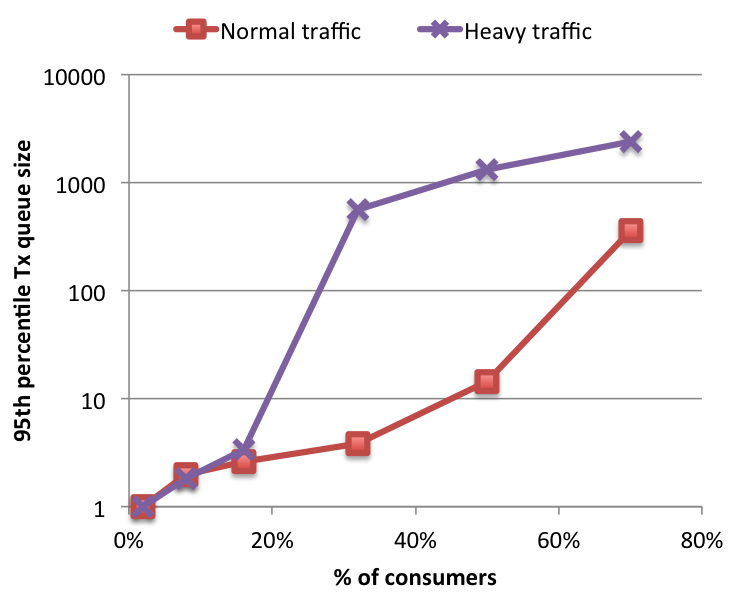}
\label{fig:TxQueueSize}
}
\end{minipage}
\vspace{-5mm}
\caption{}
\vspace{-0.5cm}
\end{figure*}

Measures the effectiveness of \mbox{in-network} caching for reducing the load on the infrastructure. Concretely, this is defined as the percentage of Interests satisfied by a Data packet coming from the cache of a node (either car or RSU).
In the GPSR case only those nodes where the streaming application is running are able to act as caches.
The results are reported in Figure~\ref{fig:InfrastructureOffload}, with an additional pair of lines, labelled ``mules only'' in the legend, where we considered only other cars as potential caches (\ie{} RSU caches were excluded).
As expected, \name leads to a substantially higher cache utilization, even when only mules are considered, while GPSR does not go beyond 20\% of Interests satisfied by caches.
Note that to avoid skewing the results to our advantage, the effect of caches on re-issued Interests described for the previous metric is not considered here.



\subsection{Handling mobility of data providers}

When data is provided by moving mules, the binding between names and geographic areas may have a short life.
\name partially copes with this problem by specifying an entire geographic area as the destination, instead of an exact point in space, and then by flooding the Interest within the destination area.
More importantly, as shown in Figure~\ref{fig:MulesPerConsumer}, a consumer is able to opportunistically retrieve chunks of the same song from different moving mules.
On one hand, enabling every receiver to cache a Data packet allows the content to spread to a large number of cars; on the other hand, NDN, focusing on content rather than nodes, renders the identity of the data provider irrelevant.
These two factors combined allow the consumer to retrieve the data in an opportunistic way from nearby nodes without having to ``follow'' a specific node.

\subsection{Simulations with higher car density}

We performed the same set of simulations using a denser car mobility.
This time the total number of cars on the map was 1048, arranged as follows: 56\% on 6-lane roads, 30\% on 4-lane roads, and the remaining 14\% on 2-lane roads.
The rest of the parameters were left untouched.

The results relative to the success rate and user satisfaction, although decreased compared to the previous mobility, clearly showed that \name can perform substantially better than GPSR even with a larger number of nodes.
The percentage of infrastructure offload raised even more, due to the fact that \name can take advantage of the improved caching opportunities offered by the denser car traffic.
However, \nameNoSpace's overhead, measured in terms of number of V2V channel accesses, also increased, and reached the same level of GPSR in the scenarios with more than 32\% of consumers.

We speculated that this performance degradation was to be ascribed to a rapid worsening of the wireless channel conditions: as more and more nodes try to transmit, the network becomes congested, the chance of collisions increases and more packets are dropped due to queues filling up.

To confirm this intuition we measured the length of the transmission queues at the MAC layer on each node.
Indeed, as Figure~\ref{fig:TxQueueSize} shows, starting with 32\% of consumers the queue length increases by two orders of magnitude in the heavy vehicular traffic scenario.
By comparison, the increase is much slower with the previous mobility.
We believe that these findings satisfactorily explain the reduced protocol efficiency observed in the high density simulations.
We intend to address this limitation of \name in a future work, by investigating congestion avoidance and congestion control techniques.

Moreover it should be noted that, while GPSR packets always follow a single path, \name may experience cases of multipath, because forwarding decisions are taken at the receiver side.
For instance, cars on different roads might decide to forward the same packet if they cannot hear each other, because both of them are closer to the destination than the previous hop.
This event can increase the overhead, but at the same time it makes the protocol more reliable, thus increasing the chances of retrieving the desired content.

\section{Discussions and Final Remarks}
\label{sec:discuss}
We developed a self-learning scheme to enable effective data delivery in highly dynamic vehicular environments.
\name strategy is to learn where the Content resides and then steer Interest towards such area.
In contrast with IP-based geo-routing, which attempts to deliver packets to a specific end node, \name forwards Interests towards the area content resides in, enabling fetching from any available data carriers within the region, either producers, mules, or RSU. \name automatically learns content's geographical location and requires no location service or oracle which are typically required by traditional Geo-routing. Furthermore, while IP-based geo-routing is connectivity-dependent and uses a one-hop hello protocol to maintain the local topology, all \name traffic is related to Interest--Data transactions, \ie{} if there is no request for content, there would be no packet in the network.  Lastly, we observed that the NDN's basic breadcrumbs mechanism is resilient to mobility: the 95$^{th}$ RTT percentile for an Interest-Data transaction is less than 300ms. Vehicles do not move far in the time elapsed between an Interest and the corresponding data thus ensuring effective retrieval of Data packets. 

\name has been extensively evaluated through simulations and features low overhead and high performances for both V2V and V2I scenarios.
Our simulation setting assumes that all RSUs can listen to the packets within their vicinity of WiFi signal reachability.
We understand that the situation can be different in real WiFi deployment today, where RSUs may not be in the same SSID domain as vehicles and thus may not be able to receive/send packets with cars.  However we believe this issue is simply the artifact of today's protocol implementation, while our goal in this paper is to explore what is achievable by NDN based V2V, without the constraints of today's implementation.

\bibliography{arxiv}


\end{document}